**Title: Prognostic Value of Transfer Learning Based Features in Resectable Pancreatic Ductal Adenocarcinoma**

**Authors:**

| #  | Name                   | Affiliations |
|----|------------------------|--------------|
| 1  | Yucheng Zhang          | 1,2          |
| 2  | Edrise M. Lobo-Mueller | 3            |
| 3  | Paul Karanicolas       | 4            |
| 4  | Steven Gallinger       | 2            |
| 5  | Masoom A. Haider       | 1,2,5        |
| 6  | Farzad Khalvati        | 1,2,6        |

**Affiliations**

1: Department of Medical Imaging, University of Toronto, Toronto, ON, Canada

2: Lunenfeld-Tanenbaum Research Institute, Sinai Health System, Toronto, ON, Canada

3: Department of Radiology, Faculty of Health Sciences, McMaster University and Hamilton Health Sciences, Juravinski Hospital and Cancer Centre, Hamilton, Ontario, Canada

4: Department of Surgery, Sunnybrook Health Sciences Centre, Toronto, ON, Canada.

5: Sunnybrook Research Institute, Toronto, ON, Canada

6: Department of Mechanical and Industrial Engineering, University of Toronto, Toronto, ON, Canada

**Abstract**

Pancreatic Ductal Adenocarcinoma (PDAC) is one of the most aggressive cancers with an extremely poor prognosis. Radiomics has shown prognostic ability in multiple types of cancer including PDAC. However, the prognostic value of traditional radiomics pipelines, which are based on hand-crafted radiomic features alone is limited. Convolutional neural networks (CNNs) have been shown to outperform these feature-based models in computer vision tasks. However, training a CNN from scratch needs a large sample size which is not feasible in most medical imaging studies. As an alternative solution, CNN-based transfer learning has shown potential for achieving reasonable performance using small datasets. In this work, we developed and validated a CNN-based transfer learning approach for prognostication of PDAC patients for overall survival using two independent resectable PDAC cohorts. The proposed deep transfer learning model for prognostication of PDAC achieved the area under the receiver operating characteristic curve of 0.74, which was significantly higher than that of the traditional radiomics model (0.56) as well as a CNN model trained from scratch (0.50). These results suggest that deep transfer learning may significantly improve prognosis performance using small datasets in medical imaging.

**Introduction**

Pancreatic Ductal Adenocarcinoma (PDAC) is one of the most aggressive malignancies with poor prognosis[1–3]. Evidences suggested that surgery can improve Overall Survival (OS) in resectable PDAC cohorts[1,2]. However, the 5-year survival rate of this group of patients is still low[1]. It is important to identify prognostic factors for the resectable cohorts to help healthcare providers in making personalized treatment decisions[4]. In resectable patients, clinicopathologic factors such as tumor size, margin status at surgery, and histological tumor grade have been studied as biomarkers for prognosis[4–6]. However, many of these biomarkers can only be assessed after the surgery and the opportunity for patient-tailored neoadjuvant therapy is lost. Recently, quantitative medical imaging biomarkers have shown promising results in prognostication of the overall survival for cancer patients, providing an alternative solution[7–9].

As a rapidly developing field in medical imaging, radiomics is defined as the extraction and analysis of a large number of quantitative imaging features from medical images including CT or MRI[8,10,11]. The conventional radiomic analysis pipeline is consisted of four steps as shown in Figure 1. Following this pipeline, several radiomic features had been shown to be significantly associated with clinical outcomes including overall survival (OS) or recurrence in different cancer sites such as lung, head and neck, and pancreas[12–17]. Using these radiomic features, patients can be categorized into low-risk or high-risk groups guiding clinicians to design personalized treatment plans[4,8]. Although limited work has been done in the context of PDAC, recent studies have confirmed the potential of new quantitative imaging biomarkers for resectable PDAC prognosis[4,12].

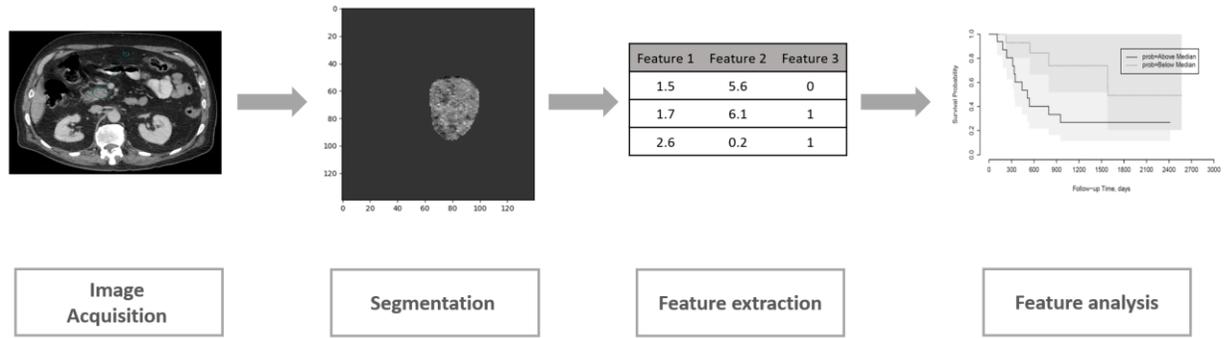

**Fig 1**. Conventional radiomics analytic pipeline

Despite the recent progress, radiomics analytics solutions have a major limitation in terms of performance. The performance of radiomics models relies on the amount of information that radiomics features can capture from medical images[8]. Most radiomics features represent morphology, first order or texture information from the regions of interest[18]. The equations of these radiomic features are often manually designed. This is a sophisticated and time-consuming process, requiring prior knowledge of image processing and tumor biology. Consequently, a poor design of the feature bank may fail to extract important information from medical images having a significant negative impact on the performance of prognostication. In contrast, deep learning models have been shown to achieve promising performances in medical imaging[19–21].

A convolutional neural network (CNN)[22,23] performs a series of convolution and pooling operations to get comprehensive quantitative information from input images[23]. Compared to hand-crafted radiomic features that are predesigned and fixed, the coefficients of CNNs are modified in the training process. Hence, the final features generated from a successfully trained CNN are "designed" to be associated with the target outcome. It has been shown that deep learning architectures are effective in different medical imaging-related tasks such as segmentation for head and neck anatomy and diagnosis for the retinal disease[24–27].

However, to train a CNN from scratch millions of parameters need to be tuned. This requires a large sample size which is not feasible in most medical imaging studies[28]. As an alternative solution, network-based transfer learning is more suitable for medical imaging-related tasks since it can achieve a comparable performance using limited amount of data[29,30].

Network-based transfer learning is defined as taking images from a different domain such as natural images (e.g., ImageNet) to build a pretrained model and then apply the pretrained model to target images (e.g., CT images of lung cancer)[31]. The idea of transfer learning is based on the assumption that the structure of a CNN is similar to the human visual cortex as both are composed of layers of neurons[30,32]. Top layers of CNNs can extract general features from images while deeper layers are able to extract information that is more specific to the outcomes[33].

Transfer learning utilizes this property, training top layers using another large dataset while finetuning deeper layers using data from the target domain. For example, the ImageNet dataset contains more than 14 million images[34]. Hence, pretraining a model using this dataset would help the model learn how to extract general features using initial layers. Given that many image recognition tasks are similar, top (shallower) layers of the pretrained network can be transferred to another CNN model. In the next step, deeper layers of the model are trained using the target domain images[35]. Since the deeper layers are more target specific, finetuning them using the images from target domain may help the model quickly adapt to the target outcome, and hence, improve overall performance.

In medical imaging, target dataset is often so small that it is impractical to properly finetune the deeper layers. Consequently, in practice, a pretrained CNN can be used as a feature extractor[13,36]. Given that convolution layers can capture high-level and informative details from images, passing the target domain images through these layers allows extractions of features. These features can be further used to train a classifier for the target domain, enabling building a high-performance transfer learning model using a small dataset.

In this study, using two independent small sample size resectable PDAC cohorts, we evaluated the prognosis performance of two different transfer learning models and compared their performance to that of the traditional radiomics feature bank. We found that transfer learning feature extraction methods provide better prognostication performance compared to conventional radiomics features, suggesting the potential of transfer learning in future medical imaging studies.

## Methods

### Dataset

Two cohorts from two independent hospitals consisting of 68 (Cohort 1) and 30 (Cohort 2) patients were enrolled in this retrospective study. All patients underwent curative intent surgical resection for PDAC from 2007 – 2012 and 2008 – 2013 in Cohort 1 and Cohort 2, respectively, and they did not receive other neo-adjuvant treatment. Preoperative portal venous phase contrast-enhanced CT images were used. Overall survival (OS) was collected as the primary outcome. To exclude the confounding effect of postoperative complications, patients who died within 90 days after the surgery were excluded. Institutional review board approval was obtained for this study from both institutions[4].

An in-house developed Region of Interest (ROI) contouring tool (ProCanVAS[37]) was used by a radiologist with 18 years of experience who completed the contours blind to the outcome (OS). Following the protocol, the slices were contoured with the largest visible cross section of the tumor on the portal venous phase. When the boundary of the tumor was not clear, it was defined by the presences of pancreatic or common bile duct cut-off and review of pancreatic phase images[4]. An example of the contour is shown in Figure 2.

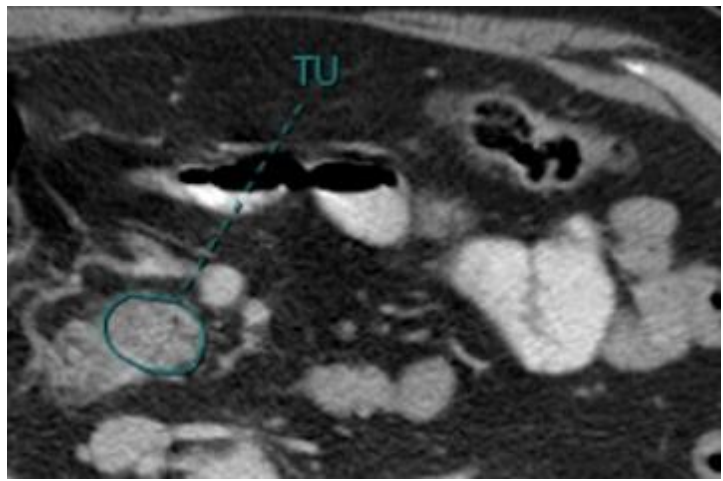

**Fig 2**. A manual contour of CT scan from a representative patient in cohort 2

**Radiomics feature extraction**

Radiomics feature was extracted using the PyRadiomics library[18] (version 2.0.0) in Python. Voxels with Hounsfield unit under -10 and above 500 were excluded so that the presence of fat and stents will not affect the feature values. The bin width (number of gray levels per bin) was set to 25. In total, 1,428 radiomic features were extracted for both cohorts. Table 1 lists different classes of features used in this study[4].

**Table 1**: List of radiomic feature classes and filters

| First-order features | Histogram-based features |
|---|---|
| Second-order texture features | Features extracted from Gray-Level Co-Occurrence matrix (GLCM) |
| Morphology features | Features based on the shape of the region of interest |
| Filters | No filter, exponential, gradient, logarithm, square, square-root, local binary pattern |

**Transfer learning**

We used two transfer learning models pretrained by non-small-cell lung carcinoma (NSCLC) CT images (LungTrans) and ImageNet images (ImgTrans) using ResNet50 architecture[34,38]. The Lung CT dataset was published on Kaggle, containing CT images from 888 lung cancer patients[39]. The ImgTrans was pretrained by ImageNet, which is an image database containing 14,197,122 images from 21,841 different categories[34]. ImageNet pretrained ResNet50 was directly available in Keras 2.0 which is a Python-based deep learning library. LungTrans ResNet50 was trained from scratch using lung CT images.

Transfer learning can be done in multiple ways depending on the sample size and the relationship between pretrained and target domains[29,32]. As shown in Figure 3, when the pretrained and target domains are similar, the features are usually extracted from the deeper layers. In contrast, when the two domains are different (e.g., natural images vs. cancer images), the features are usually extracted from the shallower layers of the pretrained network.

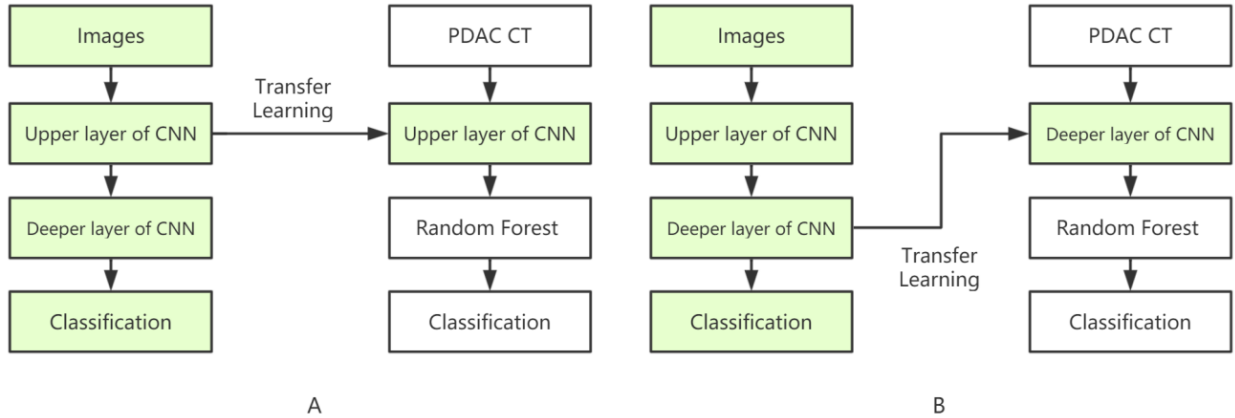

**Fig 3**. Workflow for transfer learning studies

    A. Pretrained and target domains are different.

    B. Pretrained and target domains are similar.

Given that our target domain data (PDAC CT images) was small and different from the ImageNet, with transfer learning architecture using ImgTrans, features were extracted from the $12^{th}$ layer. For LungTrans, since the domains were rather similar (CT images from lung cancer and PDAC patients), all the convolution layers were frozen, and features were extracted from the final layer. In total, 2,048 ImgTrans and 64 LungTrans features were extracted.

**Feature analysis**

Training and validation datasets were collected from two different institutions making the validation process robust and minimizing the potential overfitting. In the Cohort 1 (n=68), three prognostic models were built using features from three feature banks (PyRadiomics, ImgTrans, LungTrans) using Random Forest classifiers with 500 decision trees which is a common classifier in radiomics analytic pipeline[17].

The prognostic values of the three models were evaluated in Cohort 2 (n=30) using the area under the receiver operating characteristic (ROC) curve (AUC). Sensitivity tests were applied to test the difference between three ROC curves[40].

The predicted probabilities of death generated from the three classifiers were further treated as risk scores and tested for their prognostic power using univariate Cox Proportional Hazards Model in Cohort 2 (test set). These analyses were done in R (version 3.5.1) using "caret" , "pROC," and "survival" package[41,42].

**Results**

**Prognostic models performance**

Using features from the PyRadiomics feature bank, the Random Forest model yielded AUC of 0.56. Using ImgTrans features, the model achieved an AUC of 0.71. Finally, the AUC reached 0.74 using LungTrans features. A CNN trained on Cohort 1 from scratch with no pretraining, and then tested on Cohort 2 provided no prognostic value (AUC of ~0.50). Table 2 lists the AUC results.

**Table 2**: AUC

List of AUCs for prognostication of overall survival in the validation cohort (Cohort 2)

| Prognostic Model | AUC |
|---|---|
| **PyRadiomics** | 0.56 |
| **ImgTrans** | 0.71 |
| **LungTrans** | 0.74 |
| **CNN trained from scratch** | 0.50 |

Comparing the ROC curves using the sensitivity test[40], there was no significant difference between ROCs of PyRadiomics vs. ImgTrans and ImgTrans vs. LungTrans. Nevertheless, LungTrans feature bank had significantly higher performance than PyRadiomics feature bank with a p-value of 0.04 (AUC of 0.74 vs. 0.56). This result indicated that the transfer learning model based on lung CT images (LungTrans) significantly improved the prognostic performance of the model compared to that of the traditional radiomics methods (e.g., PyRadiomics).  Figure 4 shows the ROC curves for three models (PyRadiomics, ImgTrans, LungTrans).

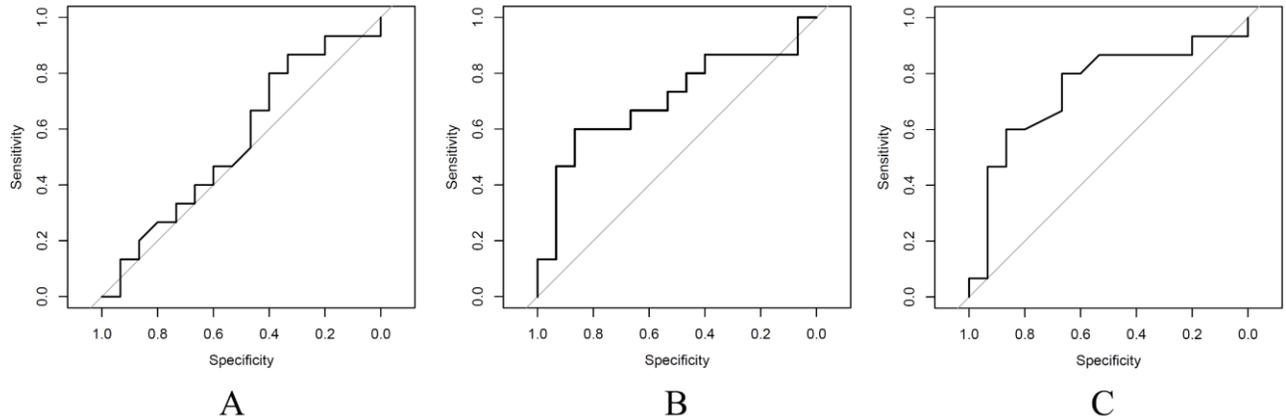

**Fig 4**. A: ROC curve using PyRadiomics feature bank only (AUC = 0.56), B: ROC curve with ImgTrans feature bank (AUC = 0.71), C: ROC curve for LungTrans feature bank (AUC = 0.74).

**Risk score**

In univariate Cox Proportional Hazards analysis, risk scores from PyRadiomics and ImgTrans prognostic models were not associated with overall survival. In contrast, the risk score from LungTrans model had significant prognostic value with p-value of 0.04. The hazard ratio (HR) and confidence intervals (CI) for risk scores generated by the PyRadiomics, ImgTrans, and LungTrans prognostic models were HR = 1.01 (Confidence Interval (CI): 0.59 – 1.76), HR = 1.24 (CI: 0.74 – 2.08), and HR = 1.78 (CI: 1.03 – 3.07), respectively, as shown in Table 3.

**Table 3**: List of hazard ratios and p values for risk scores for prognostication of overall survival in the validation cohort (Cohort 2)

| Prognostic Model | p-value | Hazard Ratio (HR) and Confidence Interval (CI) |
|:---:|:---:|:---:|
| PyRadiomic | P= 0.96 | HR = 1.01 <br><br> CI: 0.59 – 1.76 |
| ImgTrans | P = 0.42 | HR = 1.24 <br><br> CI: 0.74 – 2.08 |
| LungTrans | P = 0.04 | HR = 1.78 <br><br> CI: 1.03 – 3.07 |

Abbreviations: CI: confidence interval; ImgTrans: Deep transfer learning model pretrained by ImageNet (natural images). LungTrans: Deep transfer learning model pretrained by lung CT images.

Using the risk scores, patients can be categorized into low-risk or high-risk groups based on the median values. As shown in Kaplan-Meier plots in Figure 5, LungTrans model offers the best separation in terms of the survival patterns. This result further confirms that transfer learning feature extractor pretrained by NSCLC CT images is capable of providing prognostic information.

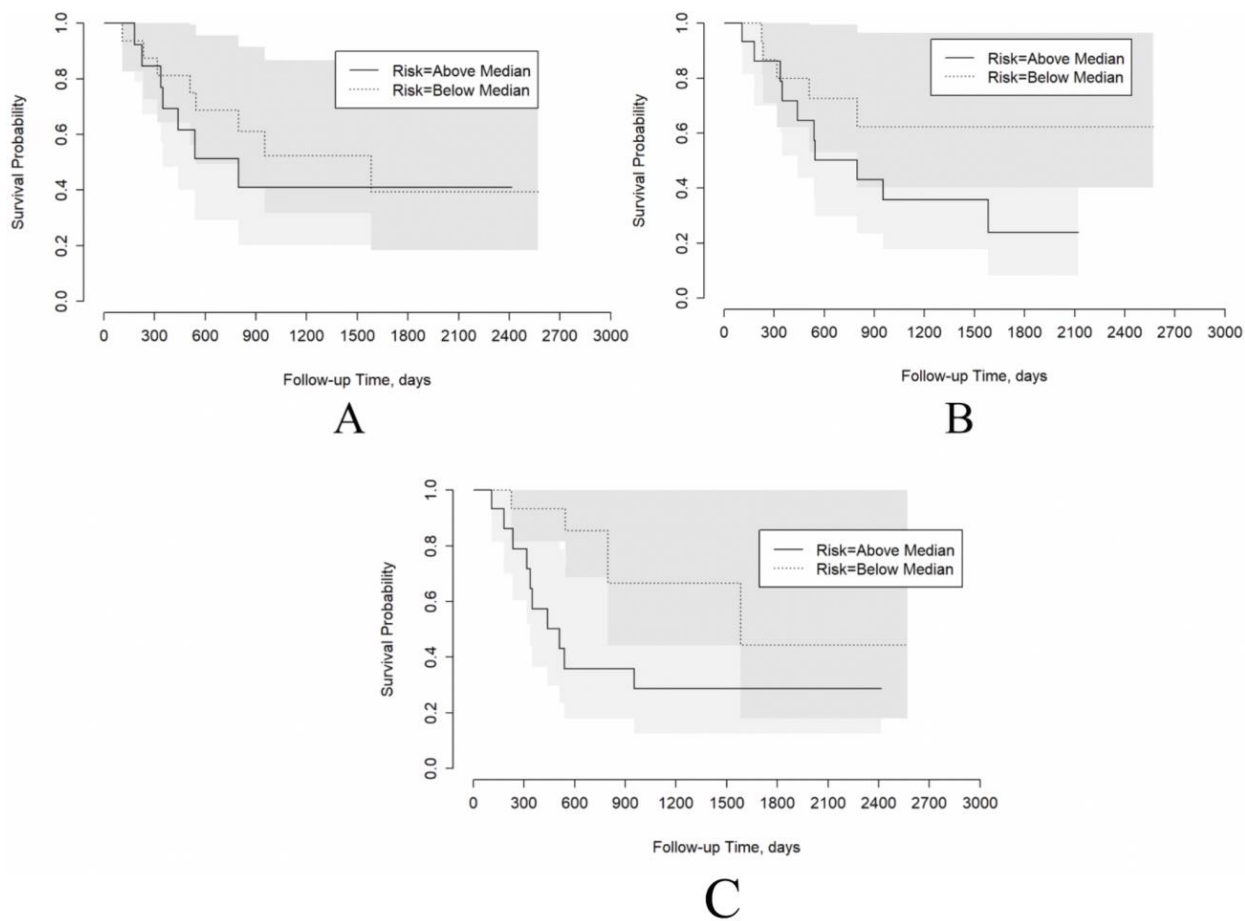

**Fig 5**. Kaplan-Meier plots for OS in Cohort 2.

    A. PyRadiomics based risk score (P=0.96)

    B. ImgTrans based risk score (P=0.42)

    C. LungTrans based risk score (P=0.04)

## Discussion

In this study, we developed and compared three prognostic models for overall survival for resectable PDAC patients using the PyRadiomics and transfer learning features banks pretrained by natural images (ImgTrans) and lung CT images (LungTrans). The LungTrans model achieved significantly better prognosis performance compared to that of the traditional radiomics approach (AUC of 0.74 vs. 0.56). This result suggested that transfer learning approach has the potential of significantly improving prognosis performance in the resectable PDAC cohort using CT images.

Previous transfer learning studies in medical imaging often utlized ImageNet pretrained models[13,29]. In our study, in addition to ImageNet (ImgTrans), we also used lung CT images to pretrain a CNN (LungTrans). These two transfer learning models provided different performances. LungTrans had higher AUC in binary survival classification and the risk score generated by LungTrans was significantly associated with overall survival. In contrast, the ImgTrans risk score was not significant. This was expected due to the substantial difference between natural images and PDAC CT images including texture, signal to noise profile and resolution.

This study showed the potential of transfer learning in a typical small sample setting. If Cohort 1 (PDAC cases alone) was used to train a CNN from scratch with no pretraining, and then tested on Cohort 2, the final output would not provide any prognostic value (AUC of ~0.50). Transfer learning, unlike conventional deep learning methods which need large datasets, can achieve reasonable performance using a limited number of samples, making it suitable for most medical imaging studies. Thus, transfer learning has the potential of playing a key role in future quantitative medical imaging studies.

Although the proposed deep transfer learning outperformed the conventional radiomics model, this is not an indication to discard radiomic features altogether. In fact, these hand-crafted features have been shown to be prognostic for survival in different cancer sites[8,43,44]. It has also been shown that feature fusion can further improve the prediction accuracy in image classification tasks[45]. Thus, an optimal feature fusion method which combines radiomic features with transfer learning features may further improve the overall performance of the prognostic model.

In this study, we aimed to improve the accuracy of survival model using the transfer learning approach. For diseases with poor prognosis, including PDAC, providing binary survival classifications offers limited information for healthcare providers for decision making since the survival rates are usually low. It would be more beneficial to provide time versus risk information, e.g., identify which time intervals had higher risks of recurrence for a resectable PDAC patient using CT images. Recent work on deep learning based survival models (e.g., DeepSurv[46]) confirm the potential for the proposed model to be integrated into a CNN-based survival model.

One limitation of the present study is the small dataset of the target domain (PDAC). A larger dataset would allow us to further investigate the effectiveness of two transfer learning approaches.

**Conclusion**

Deep transfer learning has the potential to improve the performance of prognostication for cancers with limited sample sizes such as PDAC. In this work, transfer learning models outperformed predefined radiomic model for prognostication of resectable PDAC cohorts.

**Declarations**

**Ethics approval and consent to participate**

For Cohort 1, University Health Network Research Ethics Boards approved the retrospective study and informed consent was obtained. For Cohort 2, the Sunnybrook Health Sciences Centre Research Ethics Boards approved the retrospective study and waived the requirement for informed consent.

**Authors' contributions**

YZ, MAH, and FK contributed to the design of the concept. EML, SG, MAH, and FK contributed in collecting and reviewing the data. YZ and FK contributed to the design and implementation of quantitative imaging feature extraction and machine learning modules. All authors contributed to the writing and reviewing of the paper. All authors read and approved the final manuscript. FK and MAH are co-senior authors for this manuscript.

**Competing interests**

The author(s) declare no competing interests.

**Data Availability**

The datasets of Cohort 1 and Cohort 2 analyzed during the current study are available from the corresponding author on reasonable request pending the approval of the institution(s) and trial/study investigators who contributed to the dataset.

**Funding Acknowledgment**

This study was conducted with the support of the Ontario Institute for Cancer Research (OICR, PanCuRx Translational Research Initiative) through funding provided by the Government of Ontario.